\documentclass[11pt,twoside]{article}
\usepackage{a4wide}
\newcommand{\sollsein}{\stackrel{!}{=}}
\newcommand{\aaCloseInt}{\int\hspace{-0.3cm}\int\hspace{-0.53cm}{\bigcirc}}
\newcommand{\bbbint}{\int\hspace{-0.22cm}\int\hspace{-0.22cm}\int\hspace{-0.1cm}}

\newcommand{\nl}{\newline}
\newcommand{\al}{{\alpha}}

\newcommand{\om}{{\omega}}

\newcommand{\pa}{{\partial}}
\newcommand{\beq}{\begin{equation}}
\newcommand{\eeq}{\end{equation}}
\newcommand{\beqa}{\begin{eqnarray}}
\newcommand{\eeqa}{\end{eqnarray}}
\newcommand{\ben}{\begin{enumerate}}
\newcommand{\een}{\end{enumerate}}
\newcommand{\bi}{\begin{itemize}}
\newcommand{\ei}{\end{itemize}}
\newcommand{\vr}{{\vec r}}

 \title{On the dynamics of spin systems in the Landau-Lifshitz theory}

\author{U.\ Krey\footnote{e-mail uwe.krey@physik.uni-regensburg.de }
\\
  Inst.\ f\"ur Physik II, Universit\"at Regensburg, 93040 Regensburg,
Germany
  }
\date{November 30, 2006}
\begin{document}

\large
\maketitle
\begin{abstract}

\noindent In the framework of the  Landau-Lifshitz equations without any
dissipation (an approximation which may also be helpful for finite but
weak Gilbert damping), with all interactions included (specifically
always with exchange and magnetic dipole interactions), for general
ground states, geometries and domain structures, and for many types of
effective fields (including contributions from spin currents etc.) the
{\it dynamics of the spin-precession}\, around this ground state is
considered.

 At first the precession is treated in the linear approximation. For the
(small amplitude) {\it eigenmodes}\, of the precession one has a {\it `rule
of geometric mean'}\, for the eigenfrequencies. For the eigenmodes {\it
pseudo-orthogonality relations}\, are obtained, which reflect the gyrotropic
and elliptic character of the spin precession and differ from those known
from the Schr\"odinger equation. Moreover, in contrast to quantum mechanics,
they are defined {\it locally}, not through a {\it global} average. Thus
pseudo-orthogonality statements are valid {\it everywhere} (for example
simultaneously both in the outer region and also in the core region of a
magnetic vortex line).

Then also some aspects of the nonlinear  {\it mode coupling} with
emphasis on `confluence' and `splitting' processes of elementary
magnetic spin-wave excitations are  considered. At the same time these
processes contribute to the Gilbert damping.

There are thus essential differences to quantum mechanics, although at a
first glance one discovers many similarities. From the results one may
also get insights of why these systems are so complex that (although the
essential quantities depend only on the {\it local}\, values of the
partially long-ranged effective magnetic fields) practically only
detailed experiments and computer simulations make sense.

\end{abstract}
properties) \nl {\underline{Keywords}}: Landau-Lifshitz Equs., Dynamics of
Magnetic Systems
\vglue 0.2 truecm\hrule\vglue 0.5 truecm

\section{Introduction} Recently, for possible applications in {\it
spintronics}, the dynamic behaviour of ultrathin planar magnetic
nano-structures is strongly studied ([1,2,...,9], and
and references therein). For example, one considers circular structures
with nontrivial ground states (notably with {\it vortex states}).

Theoretically, on the spatial scale of typically ten (or more) {\it
nanometers}\, (nm) and on the time scale of ten (or more) {\it
picoseconds}\, (ps) the Landau-Lifshitz equations, \cite{Landau}, with
small phenomenological `Gilbert damping', \cite{Gilbert}, is very
reliable and has found many applications in present-day simulation
software (for example, in the well-known OOMMF, \cite{Donahue}, or LLG
programs, \cite{Scheinfein}). It is `general folklore' that there are
considerable similarities between the Landau-Lifshitz-equation without
Gilbert damping and the quantum mechanical Schr\"odinger equation. This
would be advantageous, since for weak Gilbert damping one could then
exploit a lot of quantum-mechanical relations.

 Unfortunately, however, difficulties arise, since (as we will see)
 there are considerable differences from the  Schr\"odinger case.

  The present communication intends to make these points clearer; it
also serves  more general purposes, for example in the intention to go
beyond numerical simulation as far as possible.

\section{Basic theory} In the framework of the Landau-Lifshitz theory,
\cite{Landau}, the local value of the magnetic polarisation $\vec J(\vec
r, t)$ (as a function of position $\vr$ and time $t$) is described by
the following {\it ansatz}\,: \beq\vec J(\vr ,t)\equiv J_s\cdot\vec\al
(\vr ,t)\,,\eeq where $J_s$ is the saturation magnitude and $\vec\al
(\vr , t)$ the {\it direction}\, of the magnetic polarisation vector
$\vec J$. The vector $\vec\al $ is thus of unit length.

 This fact is important throughout the paper. Essentially due to this
constraint on $\vec\al (\vr ,t)$ the Landau-Lifshitz theory has the same
topological exitations (Bloch lines, Bloch point singularities, vortex
excitations, and all kinds of `topological constrictions') as in
field-theory the so-called `non-linear sigma-model', \cite{Zinn-Justin},
has; or as in elementary particle physics the corresponding
phenomenological  `Nambu and Jona-Lasinio model' has,  \cite{Nambu}.
However, since the magnetic dipole field (see below) is far-ranged, the
Landau-Lifshitz theory is in principle even more complicated than the
above-mentioned field theories (which usually do not treat interactions
corresponding to the dipole-dipole case).

 In any case, the role of the topological constraints in {\it applications}
(for example, in the wall-motion of {\it magnetic bubble-domains},
\cite{Slonczewski}, or in the  change of the polarization of the core of a
magnetic vortex-line, \cite{Waeyenberge}) should not be underestimated.

However the present paper does {\it not}\, concentrate on topological
aspects\,!

\noindent Rather, in the linear approximation (neglecting the damping
unless otherwise stated) we derive a {\it `rule of geometric mean'}\,
for the eigenfrequencies and a {\it pseudo-orthogonality relation}\, for
the eigenmodes of the precession around the nontrivial ground state. The
relations correspond to a gyrotropic and elliptic motion of the
magnetization; they differ from those known from quantum mechanics. In
particular they are defined {\it locally}.


Nonlinear terms are treated in the present paper rather
phenomenologically. We only give some schematic arguments for damping
processes based on `confluence'  and `splitting' in connection with the
so-called mode-mode coupling of excitations in vortex structures. The
results are in favour of the experiments of \cite{Buess2} and
\cite{Buess3}.

\section{Landau-Lifshitz equations} In the absence of any dissipation
(see above), on the above-mentioned spatial and temporal scales, the
equation of motion for the vector field $\vec\al (\vr ,t)$ is the
undamped Landau-Lifshitz equation, \cite{Landau}, which can be written
as \beq\label{eqLandau}\frac{{\rm d}\vec\al (\vr , t)}{{\rm d}t}
\,=\,-\gamma\cdot\vec\al (\vr , t)\times\vec H_{\rm eff}(\vr ,t)\,,\eeq
where the positive quantity $\gamma=g\cdot\frac{\mu_0 |e|}{2m_e}$ is the
so-called Land\'e factor $g$ ($\approx 2$) multiplied by the {\it
gyromagnetic ratio};  $e\,\,(=-|e|)$ is the electron's charge (a
negative quantity), $ m_e$  its mass, and $\mu_0$ the permeability of
the vacuum; $\vec H_{\rm eff}(\vr ,t)$ is an effective field, namely the
sum \beq\label{eqEffField} \vec H_{\rm eff}:=\vec H^{\rm ext}+\vec
H^{\rm dipole}+ \sum\limits_{i,k=1}^{3}\,\frac{2A_{i,k}}{J_s}\left
(\frac{\pa\vec\al}{\pa x_i}\right )\cdot \left ( \frac{\pa\vec\al}{\pa
x_k}\right ) -{\rm grad}_{\rm\vec\al}\,F_0(\vec\al )+\vec H^{\rm
therm}+\vec H^{\vec j}\,.  \eeq

In (\ref{eqEffField}), $\vec H^{\rm ext}$ is the external magnetic
field; the $A_{i,k}$ are the so-called {\it exchange constants}\, of the
crystal; $F_0(\vec\al )$ is the {\it anisotropy energy}\,, representing
the {\it easy axes}\, of the magnetization; $\vec H^{\rm therm}$ is a
fluctuating thermal field $\propto \sqrt{T}$, where $T$ is the Kelvin
temperature; this field is neglected together with the friction, to
which it is intrinsically related. Finally, $\vec H^{\vec j}$ is an
effective field proportional to the density of the electric current,
$\vec j$, or to other sources of torques acting on the magnetization.
For stationary currents all these contributions to $\vec H_{\rm eff}$ do
not depend explicitly on the time nor on the time derivative,
\cite{REMGilbert}. Often they all can be neglected with respect to the
second and/or third term on the r.h.s.\ of equation (\ref{eqEffField}).

The remaining second term on the r.h.s.\ of (\ref{eqEffField}), $\vec
H^{\rm dipole}$, is the cumbersome quantity in {\it micromagnetism},
namely the magnetic dipole field. In contrast, the third term represents
the {\it exchange interactions}, which at short wavelengths are much
stronger, such that practically they alone determine the critical
temperature $T_c$ of the ferromagnet considered. But for applications at
room temperatures and below, at distances which are much larger than the
atomic lattice constants (typically 0.2 nm) the dipole fields are most
important.

Therefore even in the $\mu m$ and $nm$ ranges, as mentioned, the
exchange interactions can often be neglected (we don't neglect them
here\,!), whereas the field $\vec H^{\rm dipole}$, which is, for
example, mainly responsible for the {\it domain structure} of a
ferromagnet, can be derived from the relation $\vec H^{\rm dipole}=-{\rm
grad}\,\Phi_m$.

The magnetostatic potential $\Phi_m$ is determined as follows:\,
\bi\item[(i)]
either  from
the {\it magnetic moment representation}
\beq\label{eqDipoldarstellung} \Phi_m(\vr
,t)\,\stackrel{(i)}{=}\,\bbbint_{\,V}\,\,{\rm d}^3x^{\,'}\,\frac{\vec
J(\vec r^{\,'} ,t)\cdot (\vr -\vec r^{\,'})} {4\pi\mu_0\,| \vr -\vec
r^{\,'}|^3}\,\eeq (here the integration volume $V$ is that of the magnetic
sample), \item[(ii)] or from the
equivalent representation of the {\it fictitious magnetic
charges}\,:
\beq\label{eqLadungsdarstellung}
\Phi_m(\vr
,t)\,\stackrel{(ii)}{=}\,\bbbint_{\,V}\,\,\,{\rm d}^3x^{\,'}\,\frac{-{\rm
div}\,\vec J(\vec
r^{\,'} ,t)} {4\pi\mu_0\,| \vr -\vec
r^{\,'}|}
+\aaCloseInt_{\,\pa V}\,\,{\rm d}^2A^{\,'}\,\frac{\vec J(\vec
r^{\,'},t)\cdot \vec n(\vec r^{\,'}) }{4\pi\mu_0\,| \vr -\vec
r^{\,'}|}\,.
\eeq
Here $\pa V$ is the (oriented) boundary of $V$; the vector $\vec n$ is
the its outer normal\,; the surface measure is ${\rm d}^2A$.
(Retardation effects do not play a role, unless the frequencies $f$
$(=\frac{\om}{2\pi})$ are larger than $\frac{c}{\Delta x}$ ($=3\cdot
10^{15}$ Hz for $\Delta x=100$ nm ).)\ei

{\it In any case, the  local value $\vec H^{\rm dipole}(\vr ,t)$ of the
dipole field depends on the  magnetization distribution of the whole
sample; i.e., in the context of the Landau-Lifshitz equation the dipole
field is a long-ranged, but local quantity, which sounds as a {\it
contradiction in itself}, but becomes nonetheless essential below}.

 Unfortunately, the magnitude of this field can also be very large,
namely at surfaces typically as large as $\frac{J_s}{\mu_0}$, which
corresponds to $\approx 2.2$ Tesla in Fe, $\approx 1.7$ Tesla in Co and
$\approx 1.0$ Tesla in permalloy. So for applications the magnetic
dipole field is as important as the exchange.

\section{Basic relations for the dynamics}
\subsection{General equations} In the following
we write $\vec \al\equiv\vec\al_0+\vec\beta$, where $\vec \al_0(\vr )$
is the static ground state, corresponding, e.g., to a non-trivial domain
structure of the sample, whereas the vector $\vec \beta (\vr ,t)$
describes the dynamics of the sample. Here we use the linear
approximation with respect to the ground state (i.e., we assume that
$|\vec\beta |$ is $\ll 1$, while $\vec\al_0(\vec r )\cdot\vec\beta (\vr
,t)\equiv 0$ and $\vec \al_0^2(\vr )\equiv 1$). As a consequence in
linear approximation also $\vec\al^2(\vr ,t)\equiv 1$, as desired.

 Furthermore we don't exclude topological singularities, e.g., Bloch
points or vortex lines, but assume that these are exclusively contained
in the {\it statics}\, of the system (i.e., in $\vec \al_0(\vr )$),
whereas $\vec\beta(\vr ,t)$ is topologically trivial (but see \cite
{Novosad}).

In the linear approximation one gets the following equation of
motion\,:
\beq \label{eq6}
  -\frac{{\rm d}\vec\beta}{{\rm d}\tilde t}=\vec\al_0 (\vr )\times \vec
  H_{\rm eff}^{\vec\beta}+\vec \beta (\vr ,\tilde t) \times \vec H_{\rm
  eff}^{\vec\al_0}\,. \eeq Here we have used the {\it reduced time}
  $\tilde t \,:=\,t\cdot \gamma $ (in the following we replace $\gamma$
  by 1,  unless otherwise stated).

But $\vec H_{\rm eff}^{\vec\al_0}(\vr )$ has the same direction as $\vec
\al^{(0)}(\vr )$ (the reader is reminded to the relation
$\vec\al^{(0)}\times \vec H_{\rm eff}=0$). Therefore the final term in
(\ref{eq6}) can be replaced by a term with a Lagrange parameter function
$\lambda^{\vec \al_0}(\vr )$\,:
\beq\label{eqMotion} -\frac{{\rm d}\vec\beta}{{\rm
d}t}=\vec\al_0 (\vr )\times\left (\vec H_{\rm eff}^{\vec\beta}(\vr
,t)-\lambda^{\vec \al_0}(\vr)\vec \beta (\vr ,\tilde t)\right ) \,. \eeq
This is a {\it linear}\, relation in $\vec\beta$. Generally only the
{\it local}\, and {\it instantaneous}\, values  of the effective field
$\vec H_{\rm eff}^{\vec\beta}$ (and of $\vec\beta$) are involved. It
should also be noted that (\ref{eqMotion}) does not explicitly depend on
$t$, see (\ref{eqDipoldarstellung}) or (\ref{eqLadungsdarstellung}).)

Now we assume the existence of {\it local}\, cartesian coordinates such
that the local $3$-axis is $\vec\al_0(\vr )$, whereas $\vec e_1(\vr )$
and $\vec e_2(\vr )$ are perpendicular to $\vec \al_0(\vr )$ and to each
other, but otherwise not yet fixed\,: Then in linear approximation the
following equation for the vector $\vec\beta(\vr ,t)$ (which also must
be perpendicular to $\vec\al_0$) results\,: \beq\label{eqBeta}
-\frac{{\rm d}}{{\rm d}t}\left (\beta_1(\vr ,t)\,\vec e_1(\vr
)+\beta_2(\vr ,t)\,\vec e_2(\vr)\right
)\,=\,\stackrel{\leftrightarrow}{\bf K} (\vr )\,\left (\beta_1(\vr
t)\,\vec e_1(\vr )+\beta_2(\vr ,t)\,\vec e_2(\vr)\right )\,.\eeq Here
$\stackrel{\leftrightarrow}{\bf K}(\vr )$ is an {\it antisymmetric}\,
real $2\times 2$-matrix, i.e., with imaginary eigenvalues (see below),
such that (after suitable rotation of the axes perpendicular to
$\vec\al_0(\vr )$) the equations for $\beta_1(\vr ,t)$ and $\beta_2(\vr
,t)$ become of the simple form \beq\label{eq8} -\frac{{\rm
d}\beta_1}{{\rm d}t} =+h_2^{eff}(\vr )\,\beta_2\eeq and \beq\label{eq9}
-\frac{{\rm d}\beta_2}{{\rm d}t} =-h_1^{eff}(\vr )\,\beta_1\,\,,\eeq
where $h_1^{eff}(\vr )$ and $h_2^{eff}(\vr )$ are {\it extremal}\,
off-diagonal values of the $2\times 2$-matrix
$\stackrel{\leftrightarrow}{\bf K} (\vr )$, see below, which we assume
to be non-negative, unless otherwise stated.

 The equations (\ref{eq8}) and (\ref{eq9}) describe an {\it elliptic
spin precession}\, around the local equilibrium spin direction $\vec
\al_0(\vr )$, and the quantities $h^{eff}_1$ and $h^{eff}_2$ are local
effective fields corresponding to the lengths of the principal axes of
the local precession ellipse.

In the following we skip the superscripts  of $h^{eff}_1$ and
$h^{eff}_2$.

\subsection{The `rule of the geometric mean' for the eigenfrequencies}

 We now make the {\it ansatz} $\beta_1(\vr ,t) =\beta_1^{(0)}(\vr
)\,\cos (\om t)$ and $\beta_{2}(\vr ,t)=\pm\beta_{2,\pm}^{(0)}(\vr
)\,\sin (\om t)$, where all quantities are {\it real}\,, $\om$ is
positive, unless otherwise stated, and both polarisations enter (the
upper (rsp.\ lower) sign in the ansatz for $\beta_2$ correspond to
mathematically positive (+) (rsp.\ negative, (-)) circularly polarised
precession; typically one has a superposition of both polarisations,
which results in the above-mentioned {\it elliptic}\, precession.
If the external field $H_3^{\rm ext}(\vr )$ and/or the exchange field
dominate the local effective field, see below, then one has purely
positive (for positive $H_3^{\rm ext}$) (rsp.\ negative, for negative
$H_3^{\rm ext}$) {\it circular}\, polarisation.).

In any case, by the {\it product}\, of (\ref{eq8}) and (\ref{eq9}) one
gets finally the eigenfrequency equation \beqa
&(\om_\nu )^2&\cdot\,\sin \om_\nu t\cdot(\pm\cos \om_\nu t)\cdot
\beta_1^{(0)}(\vr )\cdot \beta_2^{(0)}(\vr )\cr
&&\sollsein\gamma^2\cdot h_2^{(\nu
)}(\vr ,t)\cdot
h^{(\nu )}_1(\vr ,t)\cdot \cos \om_\nu t\cdot (\pm \sin \om_\nu
t)\cdot\beta_2^{(0)}(\vr )\cdot\beta_1^{(0)}(\vr )\,\,, \eeqa where the
gyromagnetic ratio $\gamma$ has been restored. For the eigenfrequencies
$\om^{\,(\nu )}$ this yields the (well-known, see
\cite{Krey})\,\,
{\it rule of the geometric mean}\, (the index $\nu$ counts the
eigenstates)\,:\beq \label{eqGeoMean}\om_\nu \equiv \gamma
\sqrt{h_1^{\,(\nu )}(\vr )\cdot h_2^{\,(\nu )}(\vr )}\,.\eeq

 Here one should note that the l.h.s.\ of the result does not depend on
  $\vr $, in contrast to the r.h.s. In fact in linear
approximation, in the Landau-Lifshitz theory, one can perform any
position-averaging of the time-dependent signal (for example, over all
spins, but usually with different weights at different positions)\,:
the eigenfrequency spectrum will always be the same. (This is useful for
applications, \cite{Buess3}, and it also allows that even from a
distant position there may be contributions to $\vec H_{\rm eff}^\beta
(\vr ,t)$.)

 The relation (\ref{eqGeoMean}) is actually at the same time more
complex and also more simple than in quantum mechanics\,: on the one
hand one has two functions, $h_1(\vr )$ and $h_2(\vr )$, instead of only
one in quantum mechanics (i.e., the potential energy $V(\vr )$), whereas
in quantum mechanics, on the other hand, not {\it all}\, space-averages,
but only the {\it full-space}\, average with the eigenfunction,
$\int\,{\rm d}^3\vr\,\psi_\nu^* (\vr )\,{\cal H}(\vr)\psi_\nu(\vr )$,
corresponds to $\hbar\om_\nu$ (in the last two equations all quantities
have their usual meaning).


If we multiply the canonical representation of the $2\times 2$-matrix
$\stackrel{\leftrightarrow}{\bf K}(\vr )$,
\beq\label{eqK}\stackrel{\leftrightarrow}{\bf K} (\vr )=
(\matrix{0&,& h_2(\vr )\cr -h_1(\vr ) &,&0}  )\,,\eeq 
\centerline{(where the
1- and 2-directions are perpendicular to $\vec \al^{(0)}(\vr )$ and
extremal (see below))}\newline by the imaginary number
$\frac{1}{\rm i}$, then equation (\ref{eqBeta}) becomes similar to a
Schr\"odinger equation with two-component vectors, (e.g., with
spin-orbit interaction), although in our case the quantities
$\frac{h_j(\vr )}{\rm i}$ (for $j=1,2$) are {\it independent}\,
imaginary functions, not necessarily fulfilling the identity
$(\frac{h_2(\vr )}{\rm i})^* =-\frac{h_1(\vr )}{\rm i}$\, (i.e., {\it
hermiticity})\,: only for $h_2\equiv h_1$ ({\it circular precession})
one would obtain $\frac{\stackrel{\leftrightarrow}{\bf K}_\nu (\vr )
}{\rm i}=\hat\sigma_y\cdot h_2(\vr )$ (i.e., a $2\times 2$-Schr\"odinger
equation involving the (hermitian) Pauli matrix $\hat\sigma_y =\left
(\matrix{0&,&-i\cr +i&,&0}\right )\,$). Instead, for {\it elliptic
precession}\, one obtains $\frac{\stackrel{\leftrightarrow}{\bf
K_\nu}(\vr ) }{\rm i}\equiv\hat\sigma_y\cdot \frac{h_2(\vr )+h_1(\vr
)}{2}+\hat\sigma_x\cdot\frac{h_1(\vr )-h_2(\vr )}{2}\cdot{\rm i}$\,\,,
with $\hat\sigma_x =\left (\matrix{0&,&1\cr 1&,&0}\right )\,$. Thus in
the elliptic case, the matrix $\frac{\stackrel{\leftrightarrow}K(\vr )
}{\rm i}$ is no longer hermitian, but still {\it
pseudo-hermitian}\,\footnote{It can be shown that this is equivalent to
the statement that the diagonalisation can only be performed with a
non-trivial Bogoliubov-Valatin transformation, see \cite{Krey}.}, in the
sense of the just-mentioned identity; see \cite{REM1}. The hermiticity
would be lost from the beginning, if the Gilbert damping were included


\subsection{Pseudo-orthogonality of the eigenmodes}


Now we complexify the eigenmodes $\vec{\beta}_\nu$ as follows\,:
Starting from $\vec\beta^{(\nu)}(\vr ,t)\equiv\beta_1^{(0)}(\vr ) \cdot
\cos (\om_\nu t)+\beta_2^{(0)}(\vr )\cdot \sin (\om_\nu t)$ we write
$\cos (\om_\nu t)={\cal R}e\, (e^{-{\rm i}\om_\nu t})$ and $\sin (\om_\nu
t)={\cal R}e\,(e^{-{\rm i}\om_\nu t}\rm i)$. In this way (by
omitting the {\it real part} ${\cal R}e$) one obtains the {\it primed}\,
quantity $\vec\beta^{(\nu )\,'}(\vr ,t):=\beta_1^{(0)}(\vr )\cdot
\,e^{-{\rm i}\om_\nu t}+\beta_2^{(0)}(\vr )\cdot{e^{-{\rm i}\om_\nu
t}}{\rm i}$\,\, (i.e., $\propto e^{-{\rm i}\om_\nu t}$). By expansion
with the complete set of eigenmodes this can be extended to general
vectors $\vec\beta (\vr ,t)$. One also sees by direct inspection that an
eigenvector ${\vec\beta}^{\,'}\propto e^{-{\rm i}\om t}$ of the equation
$-{\rm d}\vec\beta^{\,'}/{\rm d}t=\stackrel{\leftrightarrow}{\bf
K}\vec\beta^{\,'}$ with the eigenfrequency $\om_\nu$ is obtained from
the compatibility relation ${\beta}_2^{\,'} ={\rm
i}h_2^{-1/2}h_1^{1/2}{\beta}_1^{\,'}$.

In the Landau-Lifshitz formalism (denoted by the symbol $LL$) the {\it
pseudo-scalar product} of two vectors $\vec \beta$ can now be defined as
follows (i.e., {\it locally}, and with the primed quantities on the
r.h.s.)\,:
\beq\label{eqSkalarprodukt}\langle \vec\beta^{(\nu )}(\vr
\,t)|\vec\beta^{(\mu )}(\vr ,t)\rangle_{LL}\,:=\,\beta_1^{(\nu )\,'}(\vr
,t)^*\cdot h_2(\vr )\,\beta_2^{(\mu )\,'}(\vr ,t)
-\beta_2^{(\nu )\,'}(\vr ,t)^*
\cdot h_1(\vr )\,\beta_1^{(\mu )\,'}(\vr ,t)\,,\eeq 
where the *-symbol denotes the hermitian-conjugate quantity.

In this way one gets
$\vec\beta_\nu^*\propto e^{+{\rm i}\om_\nu t}$, and with (\ref{eqK})
\beq \langle \vec
\beta^{(\nu )}(\vr t)|\vec\beta^{(\mu )}(\vr
\,t)\rangle_{LL}\,:=\, {\vec\beta^{(\nu)\,'}(\vr
\,,t)}^* \cdot (\stackrel{\leftrightarrow}{\bf K}(\vec r)\,\vec\beta^{(\mu
 )\,'}(\vr ,t))\,.\eeq
This generalizes a relation of W.\ Fuller Brown, Jr., \cite{Brown}, which was
restricted to homogeneous ground-states.


\vglue 0.4 truecm \noindent The  {\it main
result}\, of the present formalism is of course the following
statement\,:\vglue 0.15 cm

{\it Two eigenmodes $\vec\beta^{(\nu )}$ rsp.\ $\vec\beta^{(\mu )}$ with
 different eigenfrequencies, $\om^{(\mu )}\ne\om^{(\nu )}$, are
 everywhere pseudo-orthogonal}.\vglue 0.15 truecm 

E.g., the so-called radial and  azimuthal eigenmodes, respectively, of
\cite{Buess1} are everywhere pseudo-orthogonal, both in the outer region and
also in the core region of a magnetic vortex line; although in the outer
region
$\vec\al=\pm\vec{e}_\varphi$, $\beta_1=\pm\beta_z$, $\beta_2=\beta_r$,
whereas in the core region $\vec\al\approx\pm\vec{e}_z$,
$\beta_1\approx\pm\beta_r$, $\beta_2=\beta_\varphi$.

 This pseudo-orthogonality, which reflects the gyrotropic and elliptic
 motion of the spins, is also valid for any spatial average, see
 \cite{Neudecker}.



Similarly to the formalism of quantum mechanics, but with the additional
freedom of locality, one  should thus be able to estimate the
eigenfrequencies from variational calculations, for example by
exploiting equation (\ref{eqGeoMean}) for  points with suitable
properties.

The proof of the above-mentioned {\it main result}\, is straightforward
(here the superscript $\bf t$ means {\it transposed}, and  it should be
noted that the primed quantities are used)\,:


\beqa
&& {\vec\beta^{(\nu)\,'}(\vr
\,,t)}^* \cdot (\stackrel{\leftrightarrow}{\bf K}(\vec r)\,\vec\beta^{(\mu
 )\,'}(\vr ,t))\equiv{-\rm i}\om_\mu\vec\beta^{(\nu )\,'\,^*}\cdot
\vec\beta^{(\mu )\,'} \cr &=& (\stackrel{\leftrightarrow}{\bf K^t}(\vr )
\,\vec\beta^{(\nu )\,'\,^*})\cdot \vec\beta^{(\mu )\,'}
=(-\stackrel{\leftrightarrow}{\bf K}(\vr ) \,\vec\beta^{(\nu
)\,'\,^*})\cdot \vec\beta^{(\mu )\,'} \equiv -{\rm
i}\om_\nu\vec\beta^{(\nu )\,'\,^*}\cdot \vec\beta^{(\mu )\,'} \,,\eeqa
i.e., one minus sign is produced by the transposition, and a second one
by the hermitian-conjugation. As a consequence, the states are
pseudo-orthogonal if the eigenfrequencies differ.


Another essential point of the result is {\it not} the above-mentioned
strange `precessional' form of the pseudo-scalar product, but rather the
fact that in contrast to quantum mechanics, which is nonlocal (see also
\cite{remNonlocal}), here the {\it local values}\, of classical
effective quantities count.

We can again apply the pseudo-orthogonality relations to vortex states. Here,
sufficiently far from the vortex core, one has $\vec{\al}(\vr ,t)\propto
\vec e_\varphi$, as already mentioned. For $h_1(\vr )$ and $h_2(\vr )$ one
should see the remarks below.

\section{Further remarks on the relation to quantum
mechanics}\label{secQuantum} Some statements on the relation to quantum
mechanics have already been presented. The following is more
subtle and at the same time more basic\,: the Landau-Lifshitz equations
form a nonlinear set of classical integro-differential equations. The
final quantities $\beta_1(\vr ,t)$ and $\beta_2(\vr ,t)$ are {\it
real}\, (only if one wants to introduce the above-mentioned
pseudo-orthogonality one is apparently forced to make the ansatz
$\propto e^{-{\rm i}\om t}$) and in linear approximation they can be
taken from a {\it real}\, Hilbert space. In contrast, in quantum
mechanics, which is totally linear, these observables are represented by
hermitian operators acting on the elements of a {\it complex}\, Hilbert
space. Although the Landau-Lifshitz equations can be derived from a
quantum mechanical model (e.g., the Heisenberg model) this involves some
kind of molecular-field approximation (e.g., products of expectation
values, $\langle \hat A\rangle\langle \hat B\rangle$, instead of
expectation values of products of operators, $\langle \hat A\hat
B\rangle$). One should of course also consider metallic ferromagnets,
which in quantum mechanics are better described by an itinerant model
instead of the simpler Heisenberg one. But in any case both correspond
to the same phenomenological Landau-Lifshitz theory. In the `itinerant
case' the atomistic derivation of the phenomenological theory is more
delicate (for the exchange field the problems have been overcome by
Korenman and coworkers, \cite{Korenman}). Therefore, although it was
shown above that the equation of motion for the classical vector $\vec
\beta (\vr ,t)$ in linear approximation looks very similar to the
Sch\"odinger equation, the correspondence to the Bogoliubov-Valatin
approach (see \cite{Krey}) is actually only true if one performs a
quantum mechanical approach from which the Landau-Lifshitz equations can
be derived.

A quantum mechanical model is also the basis of the following
section.

\section{Nonlinearity\,: mode-coupling, confluence and splitting} Now
the third-order processes, which come into play for higher oscillation
amplitudes, are treated rather schematically in an approximation
corresponding in quantum mechanics to the Bogoliubov-Valatin
formalism\,: if $\hat b_i^+$ and $\hat b_i$ are the well-known creation
and destruction operators for (Bogoliubov-Valatin) quasi-particle mode
$i$, then in the next order the Hamilton operator contains a sum (or
integral) of additional terms of the form $\delta\hat{\cal
H}:=h_-\cdot\hat b_3(\hat b_1^+\hat b_2^+) +h_+\cdot (\hat b_2\hat
b_1)\hat b_3^+\,\, $, while $h_-$ and it's hermitian-conjugate $h_+$ are
complex amplitudes of physical dimension `energy'. (In the following it
is not necessary to calculate these quantities from $\vec\al (\vr ,t)$,
although just this is implicitly done in the computer simulations.)

\noindent According to Fermi's `Golden Rule' the relaxation times for
the corresponding processes derive from integrals involving the
expression  \beq (\frac{1}{\tau_{\mp}}) :=
\frac{2\pi}{\hbar}\cdot|h_{\mp }|^2\cdot\delta(E_3-E_1-E_2)\,, \eeq
where splitting and confluence processes correspond to the $-$ or $+$
cases, respectively. (In principle this is also  not new, see
\cite{RadoSuhl}.) Of course one has $|h_+|^2\equiv |h_+^*|^2=|h_-|^2$.
However, the populations are different (see below).

 In the above $h_\mp$ factor and the ($E_3-...$) argument, respectively,
 one has both a {\it spatial}\, and also a {\it temporal}\, constraint
 corresponding to the mode form and frequency, symbolized by `$3\to
 1+2$' (`splitting') and `$1+2\to 3$' (`confluence'), with the energy
 corresponding to the frequency (through the usual relation $\Delta
 E=\hbar\Delta\om$). These contribute to the `Stokes' and  `Antistokes'
 populations, respectively. At low temperatures the Stokes population
 (`splitting') is frozen, since  excitations to be split do not exist.
 Therefore through the temperature dependence one can check the
 essentials of the mode-coupling theory.

\noindent In fact, in this way one can simultaneously understand parts
of the above-mentioned phenomenological generalisation of the
Landau-Lifshitz equation by Gilbert's damping term, \cite{Gilbert}.
Moreover, (only) if after integration the typical relaxation processes
do not distinguish a particular axis, the Gilbert damping is isotropic
{\it per ansatz}.


\section{Application to vortex states (radial and azimuthal modes)} In
the following we consider in more detail a flat circular disk with radius
$R$ of the order of 1 $\mu$ and thickness $t_h \ll R$. Outside a central
region of radius ${\cal O}(\lambda_E) :={\cal O}(\sqrt{\frac{\mu_0\cdot
2A}{J_s^2}}\,)$\,\, ($\lambda_E =\sqrt{\frac{\mu_0\cdot 2A}{J_s^2}}\approx
5.7\,\,\AA$ for Permalloy) the ground state $\vec
\al_0$ of the disc is the vortex state (i.e., $\vec\al_0(\vr ) \equiv
\hat e_\varphi$, where $\hat e_\varphi$ is the azimuthal unit vector).
The precession around the vortex state is strongly {\it elliptic},
almost {\it in-plane}. The transverse fields entering the {\it rule of
the geometric mean}\, are partially known\,: The out-of-plane component
is $h_1=-\frac{J_s}{\mu_0}\cdot \left (1+{\cal O}(\frac{t_h}{R})\right
)$ (if no external field is applied). In magnitude, this is a very large
field ($\mu_0 |h_1|\approx 1$ T for permalloy) driving the spins backwards
into in-plane direction, and practically independent of the mode index
$\nu$ (but see \cite{REM2}). This is already an exceptional case\,: to
keep ${\vec \beta}(\vr ,t)$ stationary (i.e., $\om$ {\it real}), also
$h_2$ must be negative.

Moreover, in contrast to the out-of plane value $h_1$, the in-plane
component $h_2$ depends strongly on $\nu$. If the spatial pattern of the
mode corresponds to a plane wave with (in-plane) polarisation parallel
to the magnetization, i.e., $\vec k\sim \vec \alpha_0(\vr )$ (or for an
{\it azimuthal}\, excitation of a vortex configuration, $\sim\vec
e_\varphi$) then $h_2$ almost vanishes, since there are no (effective)
magnetic charges generated in the bulk (again: but see \cite{REM2}
concerning the so-called backward modes); whereas in case of an
(in-plane) variation corresponding to a propagation $\vec k\perp \vec
\al_0(\vr )$ (or for a {\it radial}\, excitation) the effective magnetic
field $h_2(\vr )$ is maximal.

As already mentioned, the experiments of \cite{Buess3} show  the existence
of {\it radial}\, and {\it azimuthal}\, modes, respectively. In agreement
with the above statements the eigenfrequencies of the azimuthal modes are
relatively low. In fact, for the lowest modes they are less than half as
high as that of the mode corresponding to the main radial peak. Thus,
energetically, there can be {\it confluence processes}\, (at low
temperatures and at room temperature) and {\it splitting processes}\, (only
at room temperature, where they are equally probable as the confluence
processes), involving for example a {\it splitting of a radial main mode
into two azimuthal modes}. (Of course this tentative suggestion of the
quenching of these processes at low temperatures has to be carefully checked
experimentally).

\section{Conclusion}  In this communication some results (mostly not
new, but nonetheless probably useful)  for the dynamical properties
contained in the {\it Landau-Lifshitz equation}\, for ferromagnetic
systems with general ground state have been demonstrated. It was shown
that the linear approximation of these equations has strong similarities
but also strong differences compared with the quantum mechanical
Schr\"odinger equation. The essential difference is \bi \item[1)]
perhaps {\it not}\, the above {\it rule of the geometric mean},\item[2)]
also {\it not}\, the complications arising from the magnetic dipole
interaction,\item[3)] but the simple fact that {\it quantum mechanics is
nonlocal}\, (e.g., the state function $|\psi\rangle$ comprises all
values $\psi(\vr )$ throughout the space), whereas the {\it
Landau-Lifshitz theory is a classical local theory}\, (i.e., the local
values $\vec H_{\rm eff}(\vr )$ of the effective fields count, although
the sources of these fields may be at a very large distance).\ei
Generally one has got relations which are more complicated than in the
quantum mechanical case. We mention (i) again the {\it `rule of the
geometric mean'}\, for the {\it eigenfrequencies}\, and (ii) a {\it
local pseudo-orthogonality}\, of the {\it eigenmodes}, reflecting the
gyrotropic and elliptic properties of the precession. Moreover, the
Gilbert damping was partially traced back to microscopic three-( or
more-)particle processes which correspond to {\it confluence} or {\it
splitting}\, of elementary magnetic excitations as seen in recent
experiments, \cite{Buess3}.

 Essential results depend on the local properties of the magnetization
structure (for example, a domain structure). This can be at the same time
both an {\it advantage}\, and also a {\it disadvantage}\, of the formalism;
i.e., in addition to the known problems of calculating the magnetic dipole
fields one sees at this place explicitly why these systems are so complex
that one can hardly avoid a comparison with detailed numerical simulations.

 \section*{Acknowledgements} The author acknowledges stimulating
 discussions and communications from C.H.\ Back, M.\ Buess, I. Neudecker
 and K.\ Perzlmaier .


\begin{thebibliography}{99}
\bibitem{REMGilbert} It was the merit of T.L.\ Gilbert, see \cite{Gilbert},
to add to the effective field a (nonhermitian but simple) contribution
$a_G\cdot\frac{\pa \vec \al}{\pa t}$, to include the damping in a
phenomenological way instead of an equivalent elder approach of Landau and
Lifshitz, which looked much more complicated. But since the `Gilbert
constant' $a_G$ is very small, e.g., $a_G=0.008$ for permalloy, we
neglect it unless otherwise stated.
\bibitem{REM1} Here one should remember the fact that the matrices
involved are applied to $2$-vectors with components of the form
$\beta_1(\vr )\,e^{-{\rm i}\,\om t}$ or $\beta_2(\vr )\,\,{\rm
i}\,e^{-{\rm i}\,\om t}$ where however at the end the {\it real part}\, is
taken, such that the ${\rm i}$-factor in front of $\sigma_x$ in the text
simply means that instead of the $\cos (\om t)$-function one is working with
$\sin (\om t)$.
\end{thebibliography}
\end{document}